# Making Black Holes in Supernovae

S. E. Woosley[1,2] & F. X. Timmes[2,3]

[1] Max-Planck Institut für Astrophysik
85740 Garching bei München, Germany

[2] Board of Studies in Astronomy and Astrophysics
University of California, Santa Cruz, Santa Cruz, CA  95064

[3] Department of Physics and Astronomy
Clemson University, Clemson, SC  29634

## Abstract

The possibility of making stellar mass black holes in supernovae that otherwise produce viable Type II and Ib supernova explosions is discussed and estimates given of their number in the Milky Way Galaxy. Observational diagnostics of stellar mass black hole formation are reviewed. While the equation of state sets the critical mass, fall back during the explosion is an equally important (and uncertain) element in determining if a black hole is formed. SN 1987A may or may not harbor a black hole, but if the critical mass for neutron stars is 1.5 - 1.6 $M_\odot$, as Brown and Bethe suggest, it probably does. Observations alone do not yet resolve the issue. Reasons for this state of ambiguity are discussed and suggestions given as to how gamma-ray and x-ray observations in the future might help.

## 1. Introduction

Gerry Brown and his good friend Hans Bethe have often voiced the opinion that a relatively soft nuclear equation of state would lead to the existence of a large number of stellar mass black holes in our Galaxy. Brown & Bethe (1994) estimated that most stars with main sequence mass in the range 18 - 30 $M_\odot$ explode, glowing brightly as Type II supernovae and returning matter to the interstellar medium, but then go into low-mass $\gtrsim$ 1.5 $M_\odot$ black holes (see also Brown, Bruenn & Wheeler 1992). The Brown & Bethe estimates gave approximately $5 \times 10^8$ low-mass black holes in the Galaxy. Brown, Weingartner & Wijers (1995) extended these discussions to include Type Ib supernovae and to offer an explanation why observational evidence for these black holes might be scarce. Namely, the binary systems where accurate mass measurements can be made are also those where substantial mass transfer occurs. This mass transfer and subsequent loss from the Wolf-Rayet star reduces the mass of the primary to the point where it can leave a neutron star remnant (Woosley, Langer, & Weaver 1995).

Brown & Bethe (1994) developed their scenario based upon the kaon condensation equation of state of dense matter (Thorsson et al. 1994). Brown (1988) calculated a maximum neutron star mass of 1.5 $M_\odot$, a value that was later refined in Bethe & Brown (1995), based upon observations and the desire that SN 1987A produce a black hole for a certain choice of stellar models, to 1.56 $M_\odot$. The possibility that a star first explodes and subsequently drops into a black hole had been suggested earlier by Wilson et al. (1986) and Woosley & Weaver (1986). They had in mind the case of a neutron star in which high entropy and neutrinos stabilize the compact object until it cools, following the launch of a successful shock, and and then collapses into a black hole. Prakash et al. (1995) subsequently showed, for their choice of equations of state, that this was possible only for a limited interval, $\Delta M \sim 0.05$ - 0.1 $M_\odot$ above the maximum mass for a stable cold neutron star. Brown & Bethe (1994) estimated however, based in part upon unpublished calculations of Lattimer & Prakash, that an additional $\sim 0.2$ $M_\odot$ could be added from the properties of the kaon condensed equation of state, so that the total window became $\Delta M = 0.25 - 0.3 M_\odot$. Chiefly this results because at high densities the matter ends up as nuclear matter, not neutron matter. The former is "softer" than the latter, and sends the core into a black hole on a thermal time scale ($\gtrsim$ 12 s; Brown & Bethe 1994). Thus a core mass of up to $\sim$ 1.8 $M_\odot$ of baryons can briefly exist before emitting almost 0.3 $M_\odot$ of neutrinos to collapse to a black hole of gravitational mass $\sim 1.5$ $M_\odot$. The progenitor of SN 1987A probably had a baryonic iron core mass in this range and so might harbor a black hole. Bethe & Brown (1995) further stated that this relatively low maximum mass for the neutron star might be the reason why no pulsars of higher mass have been found.

In this paper we discuss our own views regarding the formation of black holes in supernovae and supernova like-events. These views are not in conflict with those of Brown & Bethe but offer, in some cases, a different perspective. In particular, we briefly review: 1) recent stellar models and the iron core mass as a function of main sequence mass, metallicity, and explosion energy; 2) the role of fall back in determining whether a black hole is formed; 3) how the number of black holes varies with the critical upper limit one adopts for the neutron star mass; 4) why the collapsed remnants of stars above $\sim$ 19 $M_\odot$ are systematically larger than those below and how the apparent cut–off in pulsar masses might be a consequence of this systematic, and 5) the evidence (and lack of evidence) for a black hole in SN 1987A.

## 2. Black Holes in the Typical Supernova

We begin by asking a provocative question, one to which the answer is almost certainly "no", but worth pursuing none the less. "Is it possible that the supernovae in our Galaxy have made no black holes?". One immediately confronts well known candidates for stellar mass black holes like Cygnus X-1 and others which probably did not grow to their present mass by accretion. So maybe the premise is transparently false, but is there anything in the basic physics or observations of supernovae or that demands black hole formation in a large fraction of events?

Brown and his colleagues were strongly influenced in their estimate that stars of 18 to 30 $M_\odot$ would leave black hole remnants by knowing the iron core masses expected from stellar evolution calculations (Woosley, Langer, & Weaver 1993, 1995; Weaver & Woosley 1993). Recent calculations, especially those of Woosley & Weaver (1995; henceforth WW95) and Timmes, Woosley,

& Weaver (1996a) allow a more precise determination of the systematics of iron core mass as a function of main sequence mass and metallicity, but are not very different from what Brown & Bethe used. For single stars between 11 and 18 $M_\odot$, the final iron core masses are relatively small and relatively insensitive to metallicity. The baryonic mass of the iron core are universally in the range 1.32 to 1.62 $M_\odot$. Most are around 1.4 $M_\odot$. Stars having main sequence masses in the 8 to 11 $M_\odot$ range were not tabulated in these papers, in part, because of their minor contribution to nucleosynthesis, but they too would develop iron cores near the Chandrasekhar limit, around 1.39 $M_\odot$. Note however, one should not fall into the trap of equating iron core masses with zero entropy Chandrasekhar masses appropriate for $Y_e = 1/2$. At least 5 significant corrections must be applied (Timmes et al 1996a), the largest of which are that $Y_e$ averages about 0.45 in the iron core and that the core has finite entropy. There is partial cancelation in these effects though so that the actual iron cores are frequently, almost accidentally, near 1.4 $M_\odot$.

These *baryonic* masses are actually a little small for the observed pulsar progenitors since one must subtract the binding energy. One needs to invoke the frequent addition of $\sim 0.1$ $M_\odot$, either because the explosion does not develop precisely at the edge of the iron core (the entropy jump at the base of the oxygen shell is frequently invoked) or because of subsequent fall back after the shock has been launched. The latter will be discussed more below. Presumably there would be universal agreement that these stars, i.e., 8 - 18 $M_\odot$, can, and probably do, leave neutron star remnants.

The stars above 18 $M_\odot$ are a different story, and for a good reason. For the current nuclear rates, especially $^{12}C(\alpha,\gamma)^{16}O$ ($S_{tot}(300$ keV$) = 170$ keV-barns), inadequate carbon is produced in the core of a star above 19 $M_\odot$ for carbon (or neon) burning to ever generate net energy at the stellar center in excess of that carried away by neutrinos generated by the pair process. Consequently carbon and neon melt away in the centers of the more massive stars without ever triggering convection which lengthens the duration of carbon and neon burning. Convective shell burning occurs, but at a higher temperature that takes less time. Consequently the star losses less entropy to neutrinos during carbon and neon burning, finishes its life with a higher central entropy, and develops a less degenerate and less centrally condensed structure. Also, more massive stars naturally have more entropy to start with and this exaggerates the effect.

As a result, stars from 19 to 40 $M_\odot$ in the WW95 survey have iron core masses in the range 1.63 to 2.02 $M_\odot$. More importantly, these iron cores are surrounded by dense shells of silicon and oxygen that will almost certainly partially accrete during the explosion. But continuing to play devil's advocate, suppose the explosion mechanism is always powerful enough to expel all material external to the iron core. This may take a lot of power. In a 35 $M_\odot$ solar metallicity model for example, the binding energy of the mantle is $2.0 \times 10^{51}$ erg and one must provide an additional $2.2 \times 10^{51}$ erg of kinetic energy at infinity, $4.2 \times 10^{51}$ erg in total, to prevent fall back and force the mass cut to be at the edge of the iron core. No one has explored modern explosion mechanisms in such massive stars except Wilson et al. (1986) and those calculations need to be redone in a more modern context (two or dimensions with convection, etc.). The calculations of Wilson did show, however, that there came a point where accretion could choke the explosion mechanism and cause a failure - i.e., a large black hole (see for example, their 100 $M_\odot$ model).

But until such explosions have been recalculated, it might be possible that the explosion energy scales up with accretion in such a way as to always guarantee the ejection of all matter outside the iron core.

The gravitational mass corresponding to 2.02 $M_\odot$ of cold baryons is about 1.7 $M_\odot$, not so much larger than Bethe & Brown's (1995) estimated limit of 1.56 $M_\odot$ and the existence of black hole remnants would then depend on acceptance of the lower value.

What of the still more massive stars? Above about 35 $M_\odot$, and for all massive stars in close interacting binaries, the hydrogen envelope should be removed, by wind or mass transfer respectively, revealing a hot compact Wolf-Rayet star. These stars are observed to have large mass loss rates and may converge on a common final mass near 4 $M_\odot$ (Woosley, Langer, & Weaver 1994). As Brown et al. (1995) also point out, the remnant mass should again be small, resembling those of the 15 $M_\odot$ stars which have similar helium core structures.

At still higher masses, for helium cores bigger than about 35 $M_\odot$ (main sequence masses $\gtrsim$ 80 $M_\odot$) one encounters stars that, if they do not lose much mass, perhaps because their metallicity is small, encounter the electron-positron pair instability at oxygen ignition. For helium cores between about 35 and 50 $M_\odot$, the instability is pulsational and drives copious mass loss. Above 50 $M_\odot$ the stars explode leaving no remnant if their helium core mass is below $\sim$ 130 $M_\odot$ (Woosley 1986). Above $M_\alpha$ =130 $M_\odot$ a black hole of increasing mass results, but such stars are very rare, at least in the present universe.

So it is possible, in principle, to build a galaxy lacking stellar mass black holes providing the (poorly understood) explosion mechanism has certain properties, fall back is negligible, and the critical gravitational mass is above 1.7 $M_\odot$. Calculations of core collapse and neutrino transport (in two and three dimensions) in the more massive supernova progenitors would help to clarify this. Now we discuss what we think happens, a picture, not too different from what Brown & Bethe have proposed, except in some details.

### 3. Fall Back

A shock moving through a region of increasing $\rho r^3$ is decelerated (Bethe 1990). This also causes the material behind the shock to slow. At late times when the shock is out of sonic communication with the center, this gives rise to the "reverse shock". At earlier times the deceleration occurs by sonic pressure waves.

The sensitivity of fall back to explosion energy in otherwise successful supernovae was first noted by Woosley (1988). WW95 studied the effect in detail and found that the deceleration of the shock in the mantle of the more massive stars is severe enough, even before the reverse shock forms, to cause appreciable matter to decelerate below the escape velocity. This material falls back over a period of hours and presumably accretes on the compact remnant (see Fig. 1 thru 4 of WW95). Additional material is added by the reverse shock about a day later.

The amount of material that falls back is quite sensitive to the explosion energy. For one 35 $M_\odot$ solar metallicity model, the collapsed remnant had a final (baryonic) mass of 7.38, 3.86, and 2.02 $M_\odot$ for kinetic energies of the ejecta at infinity of 1.23, 1.88, and 2.22 $\times 10^{51}$ erg respectively.

Even in a 15 $M_\odot$ star model, about 0.1 $M_\odot$ fell back for an explosion energy of 1.22 $\times 10^{51}$ erg. The large fall back for the more massive stars is not too surprising. The gravitational binding energy external to $10^9$ cm in the presupernova 35 $M_\odot$ star is $1.59 \times 10^{51}$ erg. Obviously the shock must have at least this much energy to eject the mantle of the star. Because of hydrodynamic interaction, it takes even more to guarantee ejection of all mass exterior to the piston. So unless the explosion mechanism, which has very limited advanced knowledge of the mantle binding energy, somehow produces much greater energy in the more massive stars than it did in SN 1987A (due to a greater accretion rate during the explosion?), there comes a point where black holes must form.

It is to be emphasized that all the models calculated in WW95, including the 35 $M_\odot$ model that left a 7.4 $M_\odot$ black hole (the number needs to be adjusted for neutrino losses), were optically bright "successful" supernova. Once a strong shock starts out, it moves faster than the star can collapse and always makes it to the surface. But obviously the nucleosynthesis and the nature of the remnant are greatly affected.

## 4. Element Histories and Black Hole Cut-Off Masses

How big a star must explode in order to make the proper solar system distribution of the elements (e.g., Timmes, Woosley, & Weaver 1995)? This is a hard question in part because of the imprecise definition of the word "explode" in the present context. As just discussed, the nucleosynthesis a star ejects is determined not only by its mass (what is present in the presupernova star - with proper adjustments for mass loss, mass transfer, and rotation) - but also by the amount of fall back which is very sensitive to the exact energy of the explosion. Unless one accepts a very strict interpretation of the Russell-Vogt Theorem, the explosion energy of a star is likely to be stochastic at, say, the 50% level, and not given uniquely by its mass. Small symmetry breaking conditions include variations in metallicity, rotation rate and angular momentum distribution, and mass loss rate.

The $\alpha$-chain nuclei in the mass region from oxygen to calcium are observed and calculated to be overproduced, relative to iron, by a factor $\simeq$ 3 in metal-deficient field halo stars and globular cluster stars (Wheeler et al. 1989; Timmes et al. 1995). That is, one expects [O thru Ca/Fe] $\simeq 0.5$ dex for the $\alpha$-chain elements. Relative to oxygen then, one expects [Mg/O] and [Si/O] to be zero, i.e. magnesium and silicon are co-produced with oxygen. On the other hand, the [C/Fe] ratio in halo and disk dwarfs is observed and calculated to be roughly constant over the entire metallicity range (Wheeler et al. 1989; Timmes et al. 1995). Massive stars synthesize sufficient carbon at low metallicities to explain the observed abundance trend, but contributions from AGB stars are necessary above [Fe/H] $\simeq$ -1.0. Carbon relative to oxygen [C/O] then, should be depressed in metal-deficient dwarfs by $\simeq$ 3 and increase to solar strength at [Fe/H]=0. In essence, [C/O] is an inverted $\alpha$-chain abundance trend.

Evolution of [Mg/O], [Si/O] and [C/O] from the Timmes et al (1995) survey are shown as solid lines in Figure 1. Magnesium and silicon relative to oxygen are generally flat, as expected, but magnesium systematically underproduces solar ratios by about -0.15 dex (a factor of 0.7 on linear scales) while silicon displays variations about solar ratios of ± 0.2 dex. Carbon to oxygen shows the expected mirror imaged α-chain abundance pattern. All three elemental (solid line) histories for [Mg/O], [Si/O] and [C/O] are within the field and halo dwarf abundance determination uncertainties.

This method has of using element histories as a probe of black hole mass cut-offs has several positive aspects. Using oxygen as the basis of comparison, instead or iron, greatly reduces the effects from parameterized explosions and the iron yield. Oxygen is a better choice than iron for a Galactic chronometer since oxygen is synthesized from a single, well-defined source: hydrostatic helium burning in presupernova stars. Iron, on the other hand, has two different sources (core collapse and thermonuclear supernovae), each operating on distinctively different time scales. However, oxygen abundances in stars are more difficult to measure, and the amount of oxygen synthesized is affected by uncertainties in the $C^{12}(\alpha,\gamma)O^{16}$ nuclear reaction rate. Nonetheless, histories of C, Mg and Si normalized to oxygen has several potential advantages.

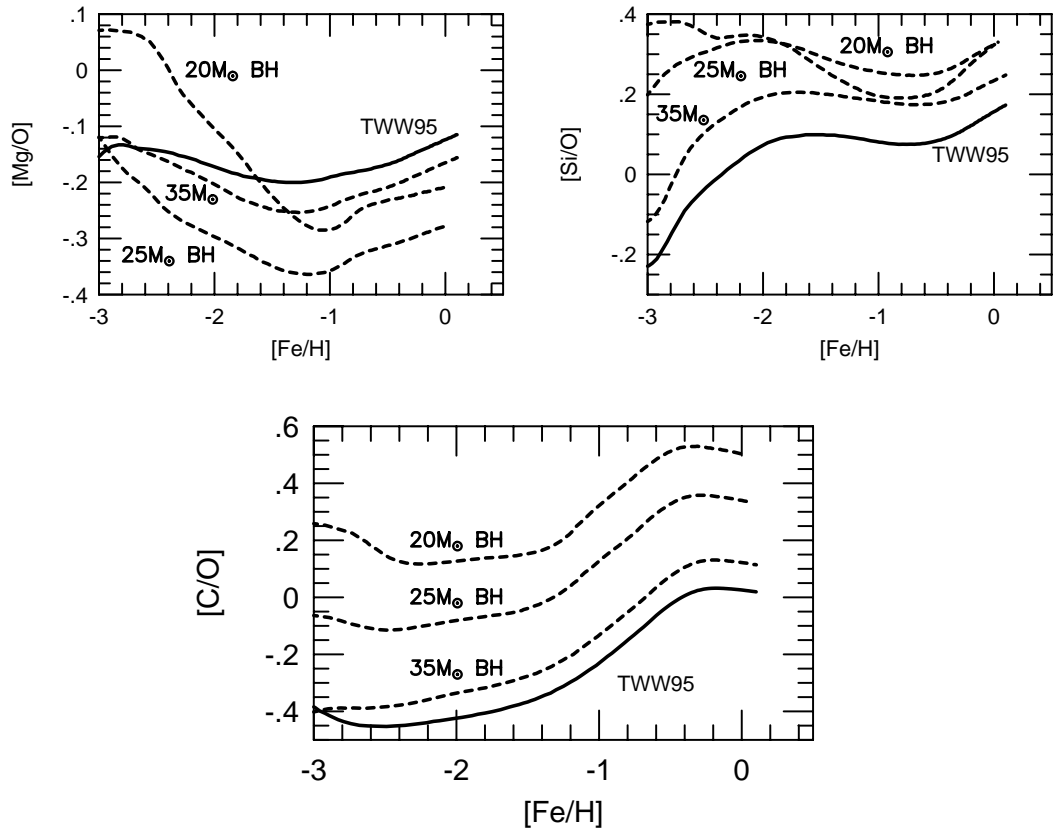

Fig. 1.— Magnesium, silicon and carbon to oxygen ratios as a function of black hole cut-off mass.

The effect of a black hole mass cut off (no yields returned to the ISM) at 35, 25 and 20 $M_\odot$ on the [Mg/O], [Si/O] and [C/O] histories are shown as dotted lines in Fig. 1. Particular care must be taken in extracting meaningful statements from these black hole mass cut–off curves. First, the time-metallicity history of each curve is not the same. Under the "no yields returned to the ISM by supernovae that make black holes" assumption, elements take different amounts of time to reach a given [Fe/H]. Second, the isotopic abundance pattern at distances and times appropriate for the presolar nebula are different for each black hole mass cut-off. Each curve does not produce an isotopic solar composition at the level attained in Figure 5 of Timmes et al. (1995). The first difficulty could be alleviated somewhat by a clever choice of normalization gauge. Calculating a set of models with different parameters chosen to assure a reasonable solar composition may help with the second problem. There are also ancillary issues of star formation rates and present epoch supernova rates becoming unacceptably large as the black hole cut–off mass is increased. It may also be likely that individual Type II or Type Ib supernova wind up with a black hole remnant and still eject a set of (reduced) yields, so that the assumption of zero chemical yield from these objects is at best a limiting case.

Most of these same caveats apply to any consideration of $\Delta Y/\Delta Z$ and black hole mass cut–off masses (e.g., Maeder 1993; Prantzos 1993; Brown & Bethe 1994). Both fall back and Wolf-Rayet winds tend to diminish the production of heavy elements with respect to helium in the more massive stars.

Noting the difficulties inherent in the interpretation of Fig. 1 and the perhaps reasonable, if arbitrary, parameterization of the WW95 survey, it appears necessary to have stars at least above 25 $M_\odot$ in order to reproduce the expected and observed [C/O], [Mg/O] and [Si/O] abundance trends. Certainly a cut–off at 20 $M_\odot$ would be intolerable. This is smaller than the 90 $M_\odot$ cut-off, depending on the adopted mass loss and $^{12}C(\alpha,\gamma)^{16}O$ rates, Prantzos (1993) suggested on the basis of the observed [C/O] ratios, Although it has seemed attractive to use element ratios and $\Delta Y/\Delta Z$ to determine a black hole mass cut-off, inconsistencies and large uncertainties in key observational and theoretical quantities (e.g., electron temperature of the H II regions and selective depletion during grain condensation) preclude strong conclusions.

## 5. The Mass Distribution of Neutron Stars and Black Holes

The upper portion of Figure 2 shows a few of the 17 neutron star binary systems favorable to measurement of their constituent masses, along with the average mass of these 17 systems (Nagase 1989; Taylor, Manchester, & Lyne 1993; Thorsett et al. 1993; Brown et al. 1995) The first 7 neutron star masses shown can be attributed to neutron star – neutron star binaries, or more generally to rotation–powered systems. Accretion–powered pulsars as a class are represented by the last 4 systems shown in Fig 2. Derived masses of accretion powered systems are more uncertain than for neutron star – neutron star binaries since Keplerian-order observations of both the pulsar's and the companion's Doppler velocity curves, not just the pulsar's, are required. Most probable masses are shown by filled circles while error bars indicate the uncertainty, with the uncertainty in PSR 1913+16 being much smaller than the filled circle.

Our calculated neutron star mass function, for the 1/2 Type II + 1/2 Type Ib case, is shown in the lower portion of Fig. 2. Other fractions for Type II and Type Ib supernovae, presupernova iron core masses as a function of the main-sequence mass, and remnant masses found after the parameterized explosion are detailed in WW95 and Timmes et al. (1996a).

Perhaps the most striking aspect of the calculated distributions is their bimodality. The birth function shown has significant peaks at 1.28 $M_\odot$ and 1.73 $M_\odot$, the amplitudes of each depending on the slope of the IMF (a Salpeter x=-1.35 was used in Fig. 2) and the the fraction ascribed to Type II events, but they are all bimodal around these two mass peaks. As discussed previously (§2), this is a consequence of the carbon abundance at the end of helium burning and whether carbon and neon burning can be exoergic at the center of the star. The width of the second peak (right hand) is certainly underestimated on its upper end because of neglect of fall back and accretion during the explosion.

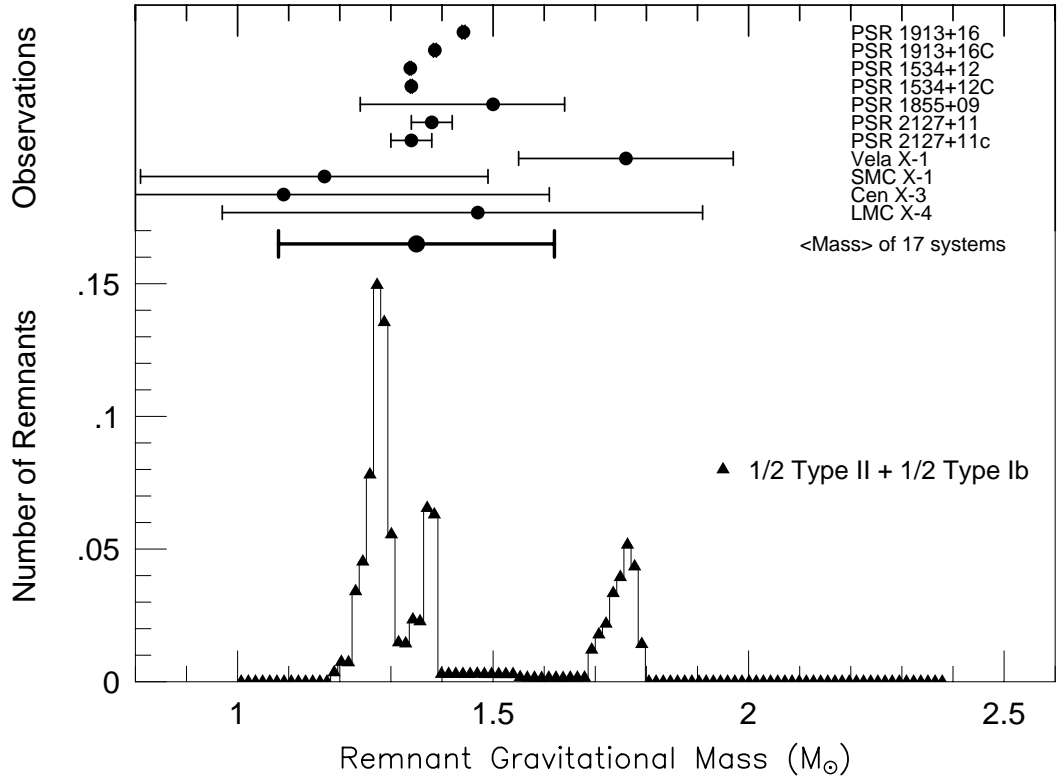

Fig. 2.— A few of the measured neutron masses and a theoretical compact remnant birth function.

The masses given in Fig. 2 are derived assuming a kinetic energy of the ejecta at infinity of 1.2 $\times 10^{51}$ erg for stars lighter than 30 $M_\odot$, and sufficient energy sufficient to eject all mass exterior to the piston above 25 $M_\odot$, with the provision that the kinetic energy of the ejecta is less than $2 \times 10^{51}$ erg. Explosion energies in $M \geq 25$ $M_\odot$ presupernova models must be increased in order to overcome the increased gravitational binding energy of the mantle (Weaver & Woosley 1993). Unless the explosion mechanism, for unknown reasons, provides a much larger characteristic energy in more massive stars, stars larger than about 30 $M_\odot$ will have a reduced set of yields and leave massive black holes. Other choices of parameters, such as the condition of constant kinetic energy at infinity, which results in a much larger amounts of falls back on high–mass stars and hence a larger number of massive remnants, are considered in Timmes et al. (1996a).

All the numbers shown in Fig. 2 are lower limits. This is because additional mass may come from sources that were either neglected in Fig. 2, or underestimated. For example, mass that accretes onto the iron core as the explosion itself is developing (the first few seconds) was not taken into account. Determining the mass additions from this source is difficult, but estimates range from 0.01 to $\sim 0.3$ $M_\odot$ for stars up to about 19 $M_\odot$, with more for the heavier stars. Mass that falls back at late times owing to hydrodynamic interaction of the shock with the mantle and envelope of the star or additional mass transfer from the binary companion could also increase the lower bounds shown in Fig 2.

The lightest neutron star produced in any of the Type II models was 1.18 $M_\odot$ and 1.22 $M_\odot$ for the Type Ib models (gravitational masses), in perhaps gratuitous agreement with the $1.2 \pm 0.26$ $M_\odot$ for 2303+46 1.2 $M_\odot$ lower limit for 1713+0747, and values for the X-ray binary systems SMC X-1, 1538-522, Cen X-3, and Her X-1 of 1.17 $M_\odot$, 1.06 $M_\odot$, 1.09$M_\odot$, and 1.04 $M_\odot$, respectively.

The integral of Fig. 2 is shown in Figure 3, with the y-axis giving the fraction of remnants with a mass greater than that remnant mass. For example, all the remnants have mass greater than 1.1 $M_\odot$ so the fraction is 1.0. For the assumed explosion parameters, no remnant had a mass greater than $\sim 1.8$ $M_\odot$ so the fraction is zero.

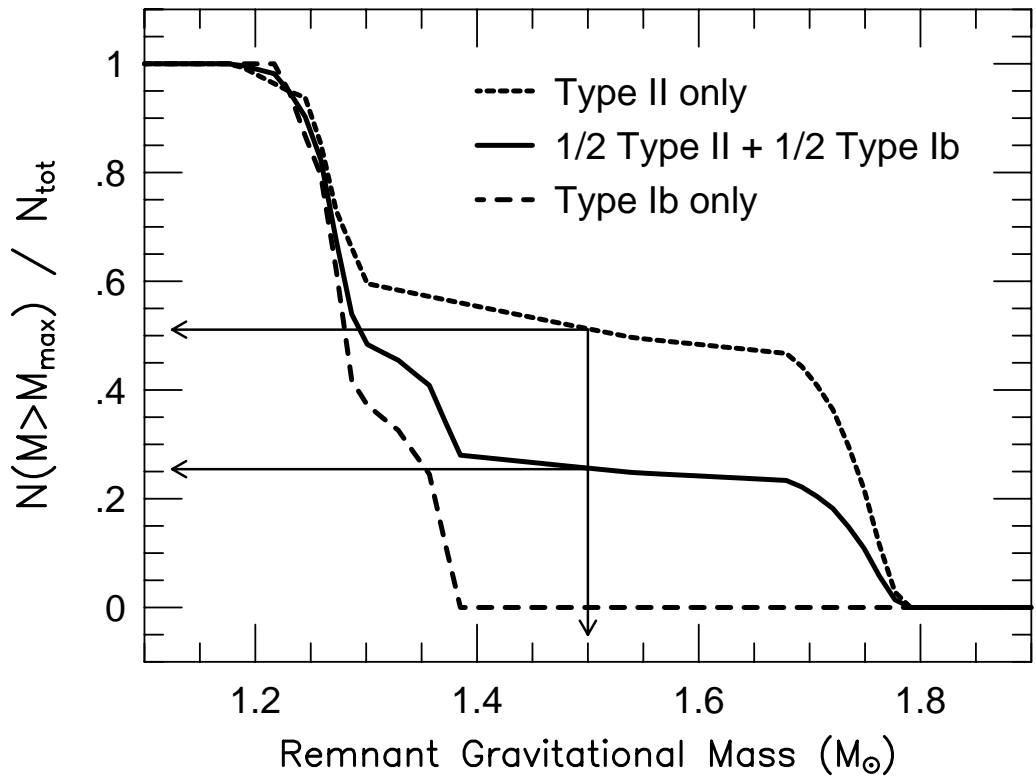

Fig. 3.— Fractions of remnants greater than a given maximum mass in the present Milky Way Galaxy.

The utility of this representation is for a given maximum neutron star mass, the fraction of remnants in the present Galaxy which are black holes and the fraction which are neutron stars may be directly read from the figure. Fig. 3 shows the example of a maximum neutron star mass of 1.5 $M_\odot$. Then 25% of all remnants in a 50-50 Type II-Ib distribution are black holes, 75% are neutron stars (all lighter than 1.5 $M_\odot$). If a purely Type II distribution is more appropriate, then 50% of all remnants are black holes with masses larger than 1.5 $M_\odot$. Converting the fractions of Fig. 3 into absolute numbers, is difficult because any estimate is dependent upon the uncertain total baryonic mass of the Galaxy (a larger reservoir has a larger number of remnants), shape of the initial mass function, integration limits of the initial mass function, and perhaps the most critical parameter, the star formation rate. Nevertheless, various estimates generally cluster around $2 \times 10^9$ supernovae explosions having occurred throughout the Galaxy's history. If the maximum neutron star mass is 1.5 $M_\odot$ and a 50-50 Type II-Ib partition is applicable, there are $5 \times 10^8$ black holes and $1.5 \times 10^9$ neutron stars in the present Galaxy.

The values shown in Fig. 3 may be underestimates because, in the volume of phase space being explored here, they assume relatively energetic explosions in stars above 25 $M_\odot$. The cut off at about 1.7 $M_\odot$ is probably not real but an artifact of neglecting additional fall back. It is interesting that there is a rapid fall off at a corresponding progenitor mass of 19 $M_\odot$, followed by a flatter tail. This is, of course, what one expects from the double peaked structure of Fig 2. The flatter tail suggests that a 1.5 $M_\odot$ maximum neutron star mass or a 1.7 $M_\odot$ maximum may not make much of a difference in the number of stellar mass black holes in the Galaxy. In nature, the tail is probably not so flat and certainly not so truncated as suggested in Fig. 3. One may make black holes of 5, 7 even 10 $M_\odot$ frequently if the explosion is weak, or begins deeper in the core than commonly cited.

## 6. Evidence for Black Hole Formation in Supernovae

What observational evidence would a Type II or Ib supernova present that it had made a black hole and not a neutron star? (SN 1987A will be treated as a special case in the next section.)

First, obviously, there would be no pulsar, and no energetic radiation from a pulsar to power a Crab-line supernova remnant. But absence of evidence is not evidence of absence. Take Cas A for example. No evidence of a neutron star has ever been observed, but still people are not certain. Perhaps the neutron star lacks an adequate rotation rate or magnetic field to be a bright pulsar. The recent observation of $^{44}$Ti lines in Cas A by COMPTEL (Iyudin et al. 1994; Schönfelder et al. 1995 Timmes et al. 1996b) might even be hard to understand if a black hole formed by fall back since $^{44}$Ti comes from the deepest layers where the shock is born. However, these observations would not rule out making a small black hole in the limited mass range discussed in §1.

One diagnostic of supernovae that make black holes because of large amounts of fall back would be the lack of ejected radioactive $^{56}$Ni. A distinguishing signature would be a bright plateau Type II supernova that plummeted to very low or zero luminosity right after the plateau - i.e., lacked the distinctive radioactive tail seen in most Type II supernovae (Patat et al. 1994). However, one would need to be careful to distinguish supernovae in the 9 - 10 $M_\odot$ range which naturally produce an anomalously small amount of $^{56}$Ni and there are other ways of producing late time emission from supernovae besides radioactivity. Spectroscopic diagnostics should help distinguish these alternatives.

## 7. A Black Hole in SN 1987A?

Brown, Bruenn & Wheeler (1992) posed the question "Did a black hole form in SN 1987A?". All agree that this did not happen during the first 10 second or so while the neutron star was still a bright source of neutrinos, but it could have collapsed minutes later as Brown & Bethe have described (see §1) or hours to days later because of fall back as WW95 have described. The appearance of many optical and $\gamma$-ray manifestations of $^{56}$Co show that fall back was not extreme, but perhaps 0.1 to 0.3 M$_\odot$ was enough. Current calculations show that this is not an unreasonable amount. If the neutron star was already hovering on the brink of instability, it would have been enough.

While considerable uncertainty still surrounds the presupernova evolution of SN 1987A, there is general agreement that, at the time of collapse it had a helium core mass of $6 \pm 1$ M$_\odot$. This corresponds to a star of about 17 to 21 M$_\odot$ on the main sequence, interestingly spanning the discontinuity in iron core masses and fall back at 19 M$_\odot$. For explosion energies around $1.2 \times 10^{51}$ erg, which is typical for SN 1987A, WW95 estimate rather large fall back masses that are, perhaps accidentally, constant at 0.3 M$_\odot$. Estimated final baryonic masses range from 1.76 to 2.02 M$_\odot$, so if an equation of state nearly as soft as Bethe & Brown suggest is correct, a black hole probably is there. But can one turn the argument around and use observational limits to prove a black hole is present and therefore limit the equation of state? We don't think so - yet.

A major point often made by the proponents of a black hole in SN 1987A is the decline of the bolometric luminosity below the Eddington value of a 1.5 M$_\odot$ neutron star for fully ionized gas, about $2 \times 10^{38}$ erg s$^{-1}$. (Of course an accreting black hole could, in principle, provide the same luminosity, but that depends upon additional assumptions regarding the angular momentum). The observed and theoretical bolometric light curves of SN 1987A are shown in Fig. 4 for the 500 to 3500 day period. The measured bolometric luminosity (e.g., Suntzeff et al. 1992) is shown as filled circles. There are no data points past day 2000 since most the emission is in the far infrared wavelength region where it is not easily observed.

Contributions from the decay of $^{56,57}$Co, $^{22}$Na, $^{44}$Ti, and $^{60}$Co are shown by the labeled dashed lines while the solid line is the total $\gamma$–ray luminosity expected from the theoretical calculations. A large, but not unphysical, ejected mass of M($^{44}$Ti) $= 1.0 \times 10^{-4}$ M$_\odot$ and M($^{60}$Co) $= 2.0 \times 10^{-5}$ M$_\odot$ was used in the theoretical light curve. Changes in the amount of $^{44}$Ti ejected produce a simple linear shift of the light curve luminosity. The labeled dashed lines indicate the $\gamma$–ray luminosity of the various radioactive isotopes and the solid line is the total luminosity, assuming that radioactive decay is the sole power source. Radioactive $^{44}$Ti tends to dominate radioactive contributions to the bolometric light curve after about 1500 days due to the half–life of $^{44}$Ti and local deposition of the positron kinetic energy ($^{44}$Sc to $^{44}$Ca). In addition, $^{60}$Co might have been contributing appreciably to the light curve at 1500 days and may contribute to the bolometric light curve at about the 10% level at 3500 days. Fig. 4 suggests that we are now entering an epoch in the supernova's life when the dominant energy source, exclusive of input from a pulsar, accreting compact object or circumstellar interaction, should be the decay of radioactive $^{44}$Ti with a small current contribution from $^{60}$Co. But there are also atomic physics effects that could give a brighter luminosity than predicted by the steady state radioactive decay (Clayton et al. 1992; Fransson & Kozma 1993) and other energy sources - a pulsar, circumstellar interaction, an embedded companion star - could also give a floor to the luminosity.

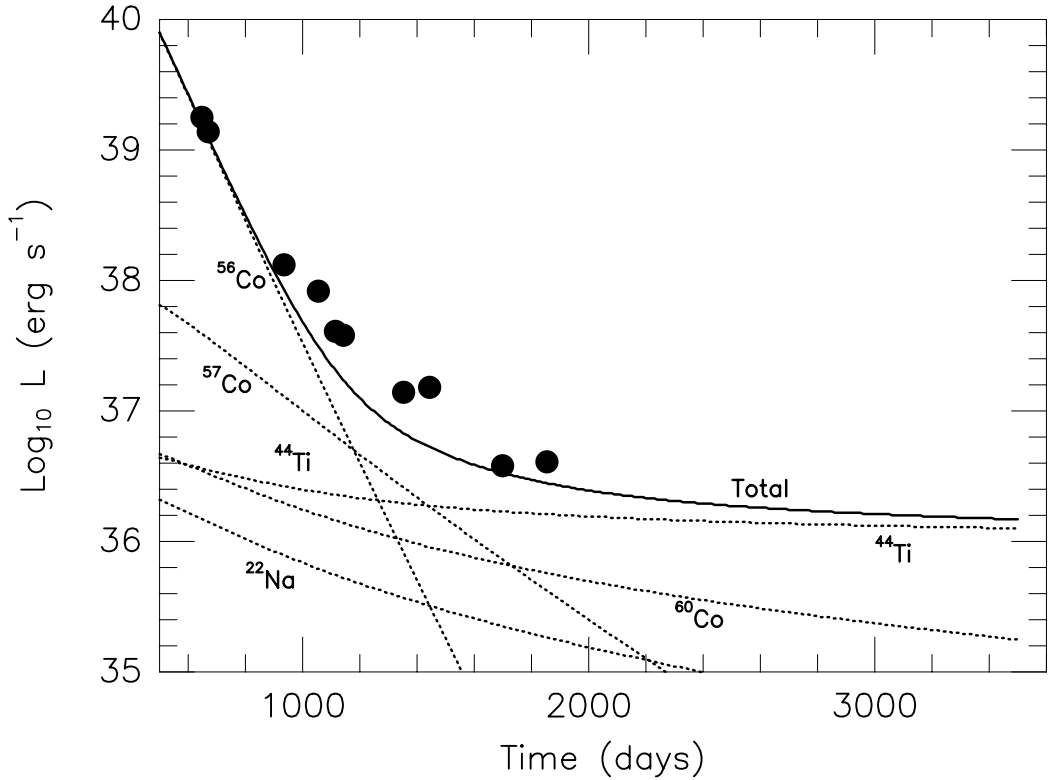

Fig. 4.— Observed bolometric light curve of SN 1987A and the expected contributions (dashed) from radioactive decay to the total (solid) bolometric light curve.

But why isn't the luminosity $10^{38}$ erg s$^{-1}$ as many expected would be the asymptotic value when SN 1987A was still bright? One possibility is that even less than $10^{-8}$ M$_\odot$ yr$^{-1}$ is finding its way to the central object. This seems like a very small value, but keep in mind that the interior of the supernova is clumpy (Li, McCray, & Sunyaev, 1993), and that the compact object may have been given a kick of many hundreds of km s$^{-1}$ (typical pulsar space velocity $\simeq 500$ km s$^{-1}$; Lyne & Lorimer 1994) and is no longer at the center of the supernova. It is also possible that the rotation of a magnetic neutron star acts to inhibit accretion that might otherwise have happened - the so called "propeller mechanism" (Shakura & Sunyaev 1976; Ghosh & Lamb 1979).

Colgate has also made the interesting observation that the pure electron scattering opacity, $\kappa_e \approx 0.2$, frequently used to estimate the Eddington luminosity may be a considerable underestimate if the accreted material is iron or even oxygen. The atomic opacity of partly ionized heavy elements could be so great as to reduce the Eddington luminosity far below $10^{38}$ erg s$^{-1}$. This suggestion deserves exploring with realistic models for the supernova composition and radiation transport.

If there is an accreting neutron star at the center and if it has a strong magnetic field it might be visible as an x-ray pulsar (Woosley, Pinto, & Hartmann 1989). The supernova is transparent to this radiation, but not to optical and radio. We don't know if there is a x-ray pulsar left over from supernova 1987A, but the recently launched X-ray Timing Explorer (XTE) satellite will address this and other fundamental issues about the nature of the remnant left by this particular Type II event. The large effective area ($\sim 0.8$ m$^2$) and broad band of sensitivity (2-200 keV) of its three instruments make it valuable for timing of intensity variations and for the determination of broad-band spectra. Even if it is now emitting pulsed radiation, debris from 1987A may be obscuring the radiation. As the debris begins to thin out, gamma and X-rays would most easily penetrate it.

One could also search for gamma-rays from the radioactive isotope, $^{44}$Ti. This is produced in the deepest layers traversed by the shock and its presence would put the tightest nucleosynthetic constraints on the mass cut (much tighter than $^{56}$Ni; but one must be cautious about mixing during the explosion). $^{44}$Ti decays to $^{44}$Sc with an accompanying emission of $\gamma$–rays at 67.85 and 78.38 keV. The $^{44}$Sc then decays to the stable isotope $^{44}$Ca with a half–life of 5.7 hr, emitting a 1.157 MeV $\gamma$–ray. The most recent measurement of the $^{44}$Ti half–life gives a preliminary value of 58 years with a 1 sigma error of 10 years (Meißner et al. 1995). For a distance of 50 kpc to SN 1987A and a half–life of 58 yr, $5 \times 10^{-5}$ M$_\odot$ of $^{44}$Ti would produce a $\gamma$–ray line flux of $\simeq 1 - 2 \times 10^{-6}$ photons cm$^{-2}$ s$^{-1}$. This line flux is too small for CGRO and probably too small for INTEGRAL instruments, but large enough that it might be detected in the next century by post-INTEGRAL experiments. Because it has greater sensitivity than COMPTEL in the 60 keV band, XTE should also search for these $^{44}$Ti lines.

This research has been supported by NSF (AST 9115367; AST 94-17161), NASA (NAGW 2525; NAG5-2843), and, in Munich, by an award from the Humboldt Foundation (SEW). FXT was supported by a Compton Gamma Ray Observatory Fellowship.

## References

Bethe, H. A. 1990, Rev. Mod.. Phys., 62, 801

Bethe, H. A., & Brown, G. E. 1995, ApJL, 445, L129

Brown, G. E. 1988, Nature, 336, 519

Brown, G. E. 1995, ApJ, 440, 270

Brown, G. E., & Bethe, H. A. 1985, Sci. Am., May, 123

Brown, G. E., & Bethe, H. A. 1994, ApJ, 423, 659

Brown, G. E., & Bethe, H. A. 1995, ApJ, 445, 129

Brown, G. E., Bruenn, S. W., & Wheeler, J. C. 1992, Comments in Astrophysics, 16, 153

Brown, G. E., & Weingartner, J. C. 1994, ApJ, 436, 843

Brown, G. E., Weingartner, J. C., & Wijers, R. A. M. J. 1995, ApJ, 463, 297

Clayton, D. D., Leising, M. D., The, L. -S., Johnson, W. N., & Kurfess, J. D. 1992, ApJ, 399, L41

Cooperstein, J. & Baron, E. 1990, in Supernovae, ed. A. Petschek (New York: Springer–Verlag), 213
Fransson, C., & Kozma, C. 1993, ApJ, 408, L25
Ghosh, P., & Lamb, F. K. 1979, ApJ, 232, 259
Iyudin, A. F., Diehl, R., Bloemen, H., Hermsen, W., Lichti, G. G., Morris, D., Ryan, J., Schönfelder, V., Steinle, H., Varendorff, M., de Vries, C., & Winkler, C. 1994, A&A, 284, L1
Li, H., McCray, R., & Sunyaev, R. A. 1993, ApJ, 419, 824
Lyne, A. G., & Lorrimer, D. R. 1994, Nature, 369, 127
Maeder, M. 1993, A&A, 268, 833
Patat, F., Barbon, R., Cappellaro, E., & Turrato, M. 1994, A&A, 282, 731
Prakash, M., Bombaci, I., M., Ellis, P. J., Lattimer, J. M., & Knorren, R. 1996, Phys. Rept., in press
Prantzos, N. 1993, A&A, 284, 477
Shakura, N. I., & Sunyaev, R. A. 1976, MNRAS, 175, 613
Schönfelder, V., et al. 1996, A&A, in press
Taylor, J. H., Manchester, R. N., & Lyne, A. G. 1993, ApJS, 88, 529
Thorsson, V., Prakash, M., & Lattimer, J. M. 1994, Nucl. Phys. A, 572, 693
Timmes, F. X., Woosley, S. E., & Weaver, T. A. 1995, ApJS, 98, 617
Timmes, F. X., Woosley, S. E., & Weaver, T. A. 1996a, ApJ, 457, 834
Timmes, F. X., Woosley, S. E., Hartmann, D. H., & Hoffman, R. D. 1996b, ApJ, in press
Weaver, T. A., & Woosley, S. E. 1993, Phys. Rept., 227, 65
Wheeler, J. C., Sneden, C., & Truran, J. W. 1989, in ARA&A Vol. 27, ed. G. Burbidge (Palo Alto: Annual Reviews), 279
Wilson, J. R., Mayle, R., Woosley, S. E., & Weaver, T. A. 1986, in Proceedings of the 12th Texas Symposium on Relativistic Astrophysics, ed. M. Livio & G. Shaviv (New York: New York Academy of Sciences), 267
Woosley, S. E. 1986, in Nucleosynthesis and Chemical Evolution 16th Advanced Course, ed. B. Hauck, A. Maeder, & G. Meynet (Swiss Society of Astrophysics and Astronomy: Geneva Observatory), 1
Woosley, S. E. 1988, ApJ, 330, 218
Woosley, S. E., & Weaver, T. A. 1986, ARAA, 24, 205
Woosley, S. E., Pinto, P. A., & Hartmann, D. H. 1989, ApJ, 346, 395
Woosley, S. E., Langer, N., & Weaver, T. A. 1993, ApJ, 411, 823
Woosley, S. E., Langer, N., & Weaver, T. A. 1995, ApJ, 448, 315
Woosley, S. E., & Weaver, T. A. (WW95) 1995, ApJS, 101, 181